\documentclass[12pt,preprint]{elsarticle}
\usepackage[dvipsnames]{xcolor}
\usepackage{amsmath}

\colorlet{FMFTred}{WildStrawberry}
\colorlet{FMFTgreen}{SeaGreen}
\colorlet{FMFTblue}{MidnightBlue}

\usepackage[colorlinks]{hyperref}

\usepackage{tikz}
\usetikzlibrary{intersections}
\usetikzlibrary{arrows,decorations.pathmorphing,backgrounds,positioning,fit,petri}
\usetikzlibrary{decorations.markings}
\usetikzlibrary{calc}

\usepackage{float}
\floatstyle{ruled}
\newfloat{workflow}{h}{lop}   
\floatname{workflow}{Diagram}

\usepackage{xspace} 

\usepackage{caption}
\usepackage[labelformat=simple]{subcaption}

\usepackage{listings}
\usepackage{graphicx}
\usepackage{tcolorbox}
\usepackage{amssymb}

\journal{Computer Physics Communications}

\newcommand{\FMFT}{\texttt{FMFT}\xspace}

\newcommand{\FORM}{\texttt{FORM}\xspace}
\newcommand{\FIRE}{\texttt{FIRE}\xspace}

\def\ep{\ensuremath{\varepsilon}}

\lstdefinelanguage{FORM}
{
  morekeywords={
    l,
    print,
    b,
    call,
    include
  },
  sensitive=false, 
  morecomment=[l]{*}, 
  morecomment=[s]{/*}{*/}, 
  morestring=[b]'' 
}

\begin{document}
    \vspace{-2cm}
\begin{frontmatter}
  \title{
    \vskip-2.5cm{\baselineskip14pt\rm
      \centerline{\normalsize DESY 17-098\hfill}
      \centerline{\normalsize July 2017\hfill}}
    \vskip1.5cm
    \texttt{\Huge \textcolor{FMFTred}{FMFT}:}  Fully Massive Four-loop Tadpoles}
    
  \author{Andrey Pikelner\corref{author}}
  
  \cortext[author] {On leave of absence from Joint Institute for Nuclear
    Research, 141980 Dubna, Russia\\
    \textit{E-mail address:} andrey.pikelner@desy.de}

  \address{II.~Institut f\"ur Theoretische Physik, Universit\"at Hamburg,\\
    Luruper Chaussee 149, 22761 Hamburg, Germany}
  
  \begin{abstract}
    We present \texttt{FMFT} --- a package written in \texttt{FORM}
    that evaluates four-loop fully massive tadpole Feynman diagrams.
    It is a successor of the \texttt{MATAD} package that has been
    successfully used to calculate many renormalization group
    functions at three-loop order in a wide range of quantum field theories especially in the Standard Model.
    We describe an internal structure of the package and provide some examples of its usage.
  \end{abstract}
  
  
\end{frontmatter}

\pagebreak

{\bf PROGRAM SUMMARY}

\begin{small}
  \noindent
  {\em Program Title:} \texttt{FMFT}                      \\
  {\em Licensing provisions:}  GPLv3                      \\
  {\em Programming language:} \texttt{FORM}               \\
  {\em Nature of problem:}\\
  Reduction of all kinds of fully massive four-loop vacuum
  integrals to the small set of master integrals.         \\
  {\em Solution method:}\\
  Reduction by means of explicit solution for IBP
  relations for topologies \textbf{X},\textbf{H} and \textbf{BMW}.
  Representation of integrals from topology \textbf{FG} as convolution
  of one-loop and two-loop massive propagator-type integrals, each
  reduced separately using generalized dimensional recurrence-relations.
  \\
  {\em Additional comments:}\\
  Available for download from URL: \url{https://github.com/apik/fmft/} \\
  \\
\end{small}

\section{Introduction}
\label{sec:introduction}

In the minimal subtraction schemes~(MS) it is usually possible to reduce
calculation of a divergent part of a given $L$-loop diagram to calculation of a
massless $L$-loop propagator-type diagram.  This is possible because
renormalization constants are independent of masses and external momenta of a
particular diagram~\cite{Collins:1974da}.  In practice, an infrared
rearrangement~(IRR) technique~\cite{Vladimirov:1979zm} is used to set all but one
external momenta and masses to zero, provided that no infrared(IR) divergences
appear.  
If there is no way to route external momentum without introducing an
IR divergences, another technique to calculate divergent parts of $L$-loop
integrals is usually used.  It is based on insertion of an equal
auxiliary mass in all
propagators of the diagram and setting all external momenta to zero, hence
reducing the problem to calculation of fully massive tadpole
integrals~\cite{Misiak:1994zw,Chetyrkin:1997fm}.

This technique was used in three-loop calculations of
Higgs self-coupling beta-function in the Standard
Model~\cite{Chetyrkin:2012rz,Chetyrkin:2013wya,Bednyakov:2013eba},
in which one needs to evaluate the divergent part of four-point Green function with
four external scalar legs. At four loops, the method was used to find anomalous dimensions
and beta-function in
QCD~\cite{vanRitbergen:1997va,Vermaseren:1997fq,Czakon:2004bu}, its
generalization for the case of extended fermion
sector~\cite{Zoller:2015tha,Chetyrkin:2016ruf} and also in the course
of calculations of higher moments of anomalous dimensions of operators of
twist-2 in QCD~\cite{Velizhanin:2014fua} and $N=4$ supersymmetric Yang-Mills theory~\cite{Velizhanin:2014zla}.
Finally, fully massive tadpole diagrams had found its application in five-loop
calculation of vacuum energy beta-function in scalar
theory~\cite{Kastening:1997ah} and recent results for five-loop QCD
renormalization constants~\cite{Luthe:2017ttc}.

All these results require calculation of thousands of fully massive
integrals. For systematic solution of such a problem the integration by
parts(IBP)~\cite{Chetyrkin:1981qh} relations are usually
applied. IBP relations allow one to reduce large number of initial
integrals to a small number of master integrals. It is possible to carry out the procedure of initial
integrals reduction to set of master integrals in an automatic way and in general case the problem can be solved by the
Laporta algorithm~\cite{Laporta:2001dd}. In some cases one can resolve recurrence relations originating from IBP identities explicitly and create a special purpose package for reduction of
integrals of special type. Famous examples of such type of solutions are
the \texttt{MINCER}~\cite{Larin:1991fz} package for reduction of three-loop massless
propagators and the \texttt{FORCER}~\cite{Ruijl:2017cxj} package, which extends 
\texttt{MINCER} to the four-loop level. The problem of reduction of
three-loop vacuum-type integrals, not necessary fully massive,
can be solved by the package
\texttt{MATAD}~\cite{Steinhauser:2000ry}. Laporta algorithm was
successfully used for reduction of four-loop tadpoles in
\cite{Schroder:2002re,Laporta:2002pg}. In this article we present the
\FMFT package for reduction of fully massive four-loop tadpoles which
can find application in calculations of four-loop renormalization
group functions in Standard Model and moments of splitting functions in QCD.

Integrals reducible by means of 
 \FMFT can be attributed to the  single auxiliary
topology~(\ref{eq:auxtopo}) with ten propagators, i.e.:

\begin{equation}
  \label{eq:auxtopo}
  I_{n_1\dots n_{10}}=\int\frac{d[k_1]d[k_2]d[k_3]d[k_4]}{D_{1}^{n_1}D_{2}^{n_2}D_{3}^{n_3}D_{4}^{n_4}D_{1;4}^{n_5}D_{2;4}^{n_6}D_{3;4}^{n_7}D_{1;2}^{n_8}D_{1;3}^{n_9}D_{1;23}^{n_{10}}},
\end{equation}
where denominators are defined as
\begin{equation}
  \label{eq:dens}
  D_a=k_a^2-m^2,\,D_{a;b}=(k_a-k_b)^2-m^2,\,D_{a;bc}=(k_a-k_b-k_c)^2-m^2.
\end{equation}
In 
\eqref{eq:dens} all masses $m^2$ are set to one during integral evaluation. For single
scale integrals the mass dependence can be easily reconstructed from
dimensional considerations. The integration measure is defined as
$d[k]=e^{\ep \gamma_E}\frac{d^d k}{i\pi^{d/2}}$.

\section{Internal structure and usage details}
\label{sec:struct-deta-oper}

\tikzset{
  sl/.style={line width=1pt},
  ndi/.style={line width=1pt,draw=FMFTblue,},
  diblue/.style={line width=1pt,draw=FMFTblue, postaction={decorate},
    decoration={markings,mark=at position .55 with {\arrow[draw=FMFTblue]{>}}}},
  dired/.style={line width=1pt,draw=FMFTred, postaction={decorate},
    decoration={markings,mark=at position .55 with {\arrow[draw=FMFTred]{>}}}},
  digreen/.style={line width=1pt,draw=FMFTgreen, postaction={decorate},
    decoration={markings,mark=at position .55 with {\arrow[draw=FMFTgreen]{>}}}}
}

\begin{figure} [h] 
  \centering
  \begin{subfigure}[b]{0.22\textwidth}
    \centering
    \begin{tikzpicture}[baseline=(bnv1.base),scale=0.35]
      \node (bnv1) at (1,0) {};
      \coordinate (v1) at (240:4);   
      \coordinate (v2) at (300:4);
      \coordinate (v3) at (60:4); 
      \coordinate (v4) at (120:4);
      \coordinate (v5) at ($(v1)!0.5!(v4)$);  
      (4pt);
      \coordinate (v6) at ($(v2)!0.5!(v3)$);  
      \draw[diblue] (v1) arc (240:120:4);  \draw (180:4) node[anchor=west] {$p_2$};
      \draw[diblue] (v2) arc (300:240:4); \draw (270:4) node[anchor=south] {$p_4$};
      \draw[diblue] (v3) arc (60:-60:4);  \draw (0:4) node[anchor=east] {$p_3$};
      \draw[diblue] (v4) arc (120:60:4); \draw (90:4) node[anchor=north]
      {$p_1$};
      \draw[diblue] (v6) -- (v5); \draw ($(v6)!0.5!(v5)$)
      node[anchor=south] {$p_5$}; 
      \draw[diblue] (v5) -- (v4); \draw ($(v5)!0.5!(v4)$)
      node[anchor=west] {$p_8$};
      \draw[diblue] (v5) -- (v1); \draw ($(v5)!0.5!(v1)$) node[anchor=west] {$p_6$};
      \draw[diblue] (v2) -- (v6); \draw ($(v2)!0.5!(v6)$)
      node[anchor=east] {$p_{7}$};
      \draw[diblue] (v3) -- (v6); \draw ($(v3)!0.5!(v6)$)
      node[anchor=east] {$p_{9}$};
      \fill (v1) circle (4pt);
      \fill (v2) circle (4pt);
      \fill (v3) circle (4pt);
      \fill (v4) circle (4pt);
      \fill (v5) circle (4pt);
      \fill (v6) circle (4pt);
    \end{tikzpicture}
    \caption{\bf H}
    \label{fig:fmft-H}
  \end{subfigure}%
  ~
  \begin{subfigure}[b]{0.22\textwidth}
    \centering
    \begin{tikzpicture}[baseline=(bnv1.base),scale=0.35]
      \node (bnv1) at (1,0) {};
      \coordinate (v1) at (240:4);   
      \coordinate (v2) at (300:4);
      \coordinate (v3) at (60:4); 
      \coordinate (v4) at (120:4);
      \coordinate (v5) at ($(v1)!0.5!(v4)$);  
      (4pt);
      \coordinate (v6) at ($(v2)!0.5!(v3)$);  
      \draw[diblue] (v4) arc (120:240:4);  \draw (180:4) node[anchor=west] {$p_2$};
      \draw[diblue] (v2) arc (300:240:4); \draw (270:4) node[anchor=south] {$p_{10}$};
      \draw[diblue] (v2) arc (-60:60:4);  \draw (0:4) node[anchor=east] {$p_3$};
      \draw[diblue] (v3) arc (60:120:4); \draw (90:4) node[anchor=north]
      {$p_4$};
      \draw[diblue] (v5) -- (v6); \draw ($(v6)!0.5!(v5)$)
      node[anchor=south] {$p_5$}; 
      \draw[diblue] (v1) -- (v5); \draw ($(v5)!0.5!(v1)$) node[anchor=west] {$p_9$};
      \draw[diblue] (v6) -- (v2); \draw ($(v2)!0.5!(v6)$)
      node[anchor=east] {$p_{8}$};
      \draw[ndi] (v6) -- ($(v6)!0.45!(v4)$); 
      \draw[diblue] ($(v6)!0.55!(v4)$) -- (v4); \draw ($(v6)!0.75!(v4)$)
      node[anchor=north east] {$p_6$}; 
      \draw[diblue] (v3) -- ($(v3)!0.5!(v5)$); 
      \draw[ndi] ($(v3)!0.5!(v5)$) -- (v5); \draw ($(v3)!0.25!(v5)$)
      node[anchor=north west] {$p_{7}$};
      \fill (v1) circle (4pt);
      \fill (v2) circle (4pt);
      \fill (v3) circle (4pt);
      \fill (v4) circle (4pt);
      \fill (v5) circle (4pt);
      \fill (v6) circle (4pt);
    \end{tikzpicture}
    \caption{\bf X}
    \label{fig:fmft-X}
  \end{subfigure}%
  %
  %
  %
  ~ 
  \begin{subfigure}[b]{0.22\textwidth}
    \centering
    \begin{tikzpicture}[baseline=(bnv1.base),scale=0.35]
      \node (bnv1) at (1,0) {};
      \coordinate (v1) at (270:4);   
      \coordinate (v2) at (0:4);
      \coordinate (v3) at (90:4); 
      \coordinate (v4) at (180:4);
      \coordinate (v5) at (0,0);  
      \draw[diblue] (v1) arc (270:180:4);  \draw (225:4) node[anchor=north
      east] {$p_5$};
      \draw[diblue] (v4) arc (180:90:4);  \draw (135:4) node[anchor=south
      east] {$p_8$};
      \draw[diblue] (v3) arc (90:0:4);  \draw (45:4) node[anchor=south
      west] {$p_3$};
      \draw[diblue] (v2) arc (0:-90:4);  \draw (-45:4) node[anchor=north west] {$p_7$};
      \draw[diblue] (v4) -- (v5); \draw ($(v4)!0.5!(v5)$)
      node[anchor=south] {$p_6$}; 
      \draw[diblue] (v3) -- (v5); \draw ($(v3)!0.5!(v5)$)
      node[anchor=west] {$p_{10}$};
      \draw[diblue] (v2) -- (v5); \draw ($(v2)!0.5!(v5)$)
      node[anchor=north] {$p_4$}; 
      \draw[diblue] (v5) -- (v1); \draw ($(v5)!0.5!(v1)$)
      node[anchor=east] {$p_{9}$};
      \fill (v1) circle (4pt);
      \fill (v2) circle (4pt);
      \fill (v3) circle (4pt);
      \fill (v4) circle (4pt);
      \fill (v5) circle (4pt);
    \end{tikzpicture}
    \caption{\bf BMW}
    \label{fig:fmft-BMW}
  \end{subfigure}
  ~
  \begin{subfigure}[b]{0.22\textwidth}
    \centering
    \begin{tikzpicture}[baseline=(bnv1.base),scale=0.35]
      \node (bnv1) at (1,0) {};
      \coordinate (v1) at (270:4);   
      \coordinate (v2) at (90:4); 
      \coordinate (v3) at (0,2);
      \coordinate (v4) at (-1.75,-1);  
      \coordinate (v5) at (1.75,-1);  
      \draw[diblue] (v1) arc (270:90:4);  \draw (180:4) node[anchor=west]
      {$p_6$}; 
      \draw[diblue] (v2) arc (90:-90:4);  \draw (0:4) node[anchor=east] {$p_8$};
      \draw[diblue] (v1) arc (240:180:3.5); \draw ($(v1)!0.5!(v4)$) node[anchor=south] {$p_1$};
      \draw[diblue] (v4) arc (180:120:3.5) ;\draw ($(v4)!0.6!(v3)$)
      node[anchor=south east] {$p_9$};
      \draw[diblue] (v5) arc (0:-60:3.5);  \draw ($(v5)!0.5!(v1)$) node[anchor=south] {$p_4$}; 
      \draw[diblue] (v5) arc (0:60:3.5);  \draw ($(v5)!0.6!(v3)$)
      node[anchor=south west] {$p_7$}; 
      \draw[diblue] (v4) -- (v5); \draw ($(v4)!0.5!(v5)$)
      node[anchor=south] {$p_3$}; 
      \draw[diblue] (v3) -- (v2); \draw ($(v3)!0.5!(v2)$)
      node[anchor=west] {$p_5$}; 
      \fill (v1) circle (4pt);
      \fill (v2) circle (4pt);
      \fill (v3) circle (4pt);
      \fill (v4) circle (4pt);
      \fill (v5) circle (4pt);
    \end{tikzpicture}
    \caption{\bf FG}
    \label{fig:fmft-FG}
  \end{subfigure}
  \caption{Base topologies with 9 lines~\subref{fig:fmft-H},\subref{fig:fmft-X},
    and 8 lines~\subref{fig:fmft-BMW},\subref{fig:fmft-FG}}\label{fig:fmft-topos}
\end{figure}

Since \FMFT is written in \FORM~\cite{Vermaseren:2000nd}, its installation reduces to
extraction of distribution archive to an appropriate place. 
For proper operation of the \FMFT package at least \FORM version 4 is
required. The latter supports \texttt{PolyRatFun}, which is used for
polynomial division and factorization.

The main steps of operation of the \FMFT package are explained in
the dia.~\ref{wflow}. 
The detailed description of each step can be found in the following sections.

\subsection{Reduction of the top-level topologies}
\label{sec:reduction-top-level}

We can associate any fully massive four-loop tadpole integral with one
of two top-level nine propagator topologies: planar \textbf{H}  
(see fig.~\ref{fig:fmft-H}) and nonplanar \textbf{X} (see
fig.~\ref{fig:fmft-X}) or their subtopologies.
If we shrink one of the lines in a diagram corresponding to topologies \textbf{H} or
\textbf{X}, we get an integral corresponding to either \textbf{BMW}
(see fig.~\ref{fig:fmft-BMW}) or \textbf{FG} (see fig.~\ref{fig:fmft-FG}).
All other integrals can be associated with topology
\textbf{FG} and its subtopologies.

As a rule, in the course of integral reduction the most time-consuming
part is not the reduction of integrals of the top-level topologies, e.g.,
\textbf{X},\textbf{H} and \textbf{BMW}, but the reduction of
integrals with smaller number of lines, which still have large powers of
propagator denominators and numerator. All such integrals in the
case of four-loop tadpoles can be mapped onto the topology \textbf{FG}. Due to this,
for topologies \textbf{X}, \textbf{H} and \textbf{BMW} we use
recurrence relations obtained from the rules generated by the 
the \texttt{LiteRed} package ~\cite{Lee:2012cn}. Our main goal, however, is to optimize the reduction of the topology \textbf{FG}.

\subsection{Reduction of topology {\normalfont\textbf{FG}}}

The main idea of the reduction strategy for topology \textbf{FG}
implemented in \FMFT is based on the observation that the
corresponding integrals can be represented as a convolution of two
propagator-type integrals:

\begin{equation}
  \label{eq:topoFGsplit}
  J_{\rm FG}= \int d[p] \left( 
      \begin{tikzpicture}[baseline=(bnv1.base),scale=0.5]
        \node (bnv1) at (0,-0.15) {}; 
        \coordinate (v1) at (-3,0); 
        \coordinate (v2) at (3,0); 
        \coordinate (v3) at (0,1.75); 
        \coordinate (v4) at (0,-1.75);
        \coordinate (v5) at (5,0); 
        \coordinate (v6) at (8,0);
        \draw[digreen] (v2) arc (-30:-90:3.5); 
        \draw (1.3,-1.4) node[anchor=south] {{\footnotesize$k_2-p$}}; 
        \draw[digreen] (v4) arc (-90:-150:3.5); 
        \draw (-1.3,-1.4) node[anchor=south] {{\footnotesize$k_1-p$}};
        \draw[digreen] (v1) arc (150:90:3.5); 
        \draw (-1.3,1.4) node[anchor=north] {{\footnotesize$k_1$}}; 
        \draw[digreen] (v3) arc (90:30:3.5); 
        \draw (1.3,1.4) node[anchor=north] {{\footnotesize$k_2$}};
        \draw[digreen] (v3) -- (v4); 
        \draw ($(v3)!0.5!(v4)$) node[anchor=west] {{\footnotesize$k_1-k_2$}};
        \draw[dired] (v6) arc (-30:-150:1.75); 
        \draw (6.5,-1) node[anchor=north] {{\footnotesize$k_4-p$}}; 
        \draw[dired] (v5) arc (150:30:1.75); 
        \draw (6.5,1) node[anchor=south] {{\footnotesize$k_4$}};
        \draw[diblue] (v2) -- (v5); 
        \draw ($(v2)!0.5!(v5)$) node[anchor=south] {{\footnotesize$p$}};
        \draw[dotted] (v6) -- (9,0);
        \draw[dotted] (-4,0) -- (v1);
        \fill (v1) circle (4pt); \fill (v2) circle (4pt); \fill (v3)
        circle (4pt); \fill (v4) circle (4pt); \fill (v5) circle (4pt);
        \fill (v6) circle (4pt);
      \end{tikzpicture}
    \right)\,.
  \end{equation}

One of these integrals is two-loop (\textbf{F}), while 
another one is one-loop (\textbf{G}). The main difference between the
standard IBP reduction and the proposed approach is the application of
reduction to each part separately. Both
parts are propagator-type diagrams with all propagators having equal
masses and arbitrary external momentum. For the reduction of one- and
two-loop propagator-type integrals with arbitrary masses and external
momenta there exists a closed-form solution as a set of generalized recurrence
relations~\cite{Tarasov:1997kx}. The latter are  also implemented in the form of
\texttt{Mathematica} package \texttt{TARCER}~\cite{Mertig:1998vk}.

\begin{workflow}
  \begin{tcolorbox}
    \begin{enumerate}
    \item Apply the reduction rules for topologies
      \textbf{X}, \textbf{H}, \textbf{BMW}
      
    \item Map the integrals onto topology \textbf{FG} expressible as a 
      convolution of two-loop (\textbf{F}) and one-loop (\textbf{G}) 
      integrals
      
    \item Reduce tensor one-loop integral corresponding to the \textbf{G}
      part of the integral to a  scalar one with shifted space-time dimension
      
    \item Reduce the \textbf{F} part of the integral with a numerator:

      \begin{enumerate}
      \item cancel scalar products, if possible
      \item substitute integrals with irreducible scalar products by scalar integrals in shifted
        space-time dimension (by means of tables)
      \item apply dimension recurrence relations to convert integrals
        with  shifted space-time dimension to the initial dimension

      \item reduce the obtained scalar integrals (in original space-time dimension)
        to a set of master integrals
      \end{enumerate}
            
    \item Do partial fractioning in $p^2$ (the momentum, external to \textbf{F} and \textbf{G})
      
    \item Rewrite $({\rm 1-loop})\otimes({\rm 2-loop})$ as an integral \textbf{FG} with a different mass on
      line with $p_5$ and arbitrary power $n_5$
      
    \item Apply recurrence relations to reduce the power $n_5$ of the topology
      \textbf{FG} to zero or one
    \end{enumerate}
  \end{tcolorbox}
  \caption{Main steps of four-loop tadpoles reduction with
    \texttt{FMFT}}
  \label{wflow}
\end{workflow}

Possible presence of numerators involving scalar products like
$k_1\cdot k_4$ or $k_2\cdot k_4$, which connect both integrals and do not let 
to apply reduction rules immediately. To disentangle integrals we
need to apply tensor reduction to one of the integrals first. The
easiest way is to express one-loop integral \textbf{G} in
the form: 

\begin{eqnarray}
  \label{eq:Gdef}
  G_{\mu_1\dots\mu_r}(n;n_1,n_2) & = & \int d^n k\frac{k_{\mu_1}\dots
                                           k_{\mu_r}}{d_1^{n_1}d_2^{n_2}},\\
  d_1 = k^2 -m^2 &,& d_2 = (k-p)^2-m^2.\nonumber
\end{eqnarray}
Then applying the general formula for one-loop tensor integral
reduction~\cite{Davydychev:1991va}, we can express it as a sum of
scalar integrals with shifted space-time dimension. For the one-loop
propagator case the general expression is reduced to

\begin{align}
  \label{eq:Gtens}
  G_{\mu_1\dots\mu_r}(d;n_1,n_2) &= (-1)^r \sum\limits_{j=0}^{[r/2]}\left(-\frac{1}{2}\right)^j\left\{[g]^j[p]^{r-2j}\right\}_{\mu_1\dots\mu_r}\\
                                 & \times
                                   \frac{\Gamma(n_1+r-2j)}{\Gamma(n_1)}G(d+2(r-j);n_1+r-2j,n_2),\nonumber 
\end{align}
where the structure
\begin{equation*}
  \label{eq:gp-sym}
  \left\{[g]^a[p]^b\right\}_{\mu_1\dots\mu_r}
\end{equation*}
is symmetric with respect to $\mu_1\dots\mu_r$ Lorentz indices and 
is constructed from $a$ metric tensors $g_{\mu\nu}$ and $b$
momenta $p$.

The resulting scalar integrals with shifted space-time dimension can be
reduced to master integrals in the initial space-time dimension by means of 
dimension recurrence relations(DRR) from~\cite{Tarasov:1997kx}. The latter are also used 
in the course of two-loop integral reduction (see below).

\subsection{Generalized recurrence relations and the reduction of two-loop massive propagator-type integrals}

After splitting the topology \textbf{FG} into parts and the tensor reduction
of one-loop subdiagram via~(\ref{eq:Gtens}), we end up with a convolution of the \emph{scalar} one-loop
propagator-type diagram and a two-loop diagram with a numerator. 
The most general form of two-loop diagram can be represented as 
\begin{equation}
  \label{eq:Tabxyz}
  T^{a b x y z}_{n_1 n_2 n_3 n_4 n_5}=\int d[k_1] d[k_2] 
  \frac{
    (k_1 \cdot p)^a (k_2 \cdot p)^b 
    (k_1 \cdot k_1)^x (k_2 \cdot k_2)^y
    (k_1 \cdot k_2)^z}
  {C_1^{n_1} C_2^{n_2} C_3^{n_3} C_4^{n_4} C_5^{n_5}}\, ,  
\end{equation}
where massive denominators are introduced in accordance with~(\ref{eq:topoFGsplit}):
\begin{align}
  \label{eq:tarcer-dens}
  C_1=k_1^2-m^2, \, C_2=k_2^2-m^2,\, & C_3=(k_1-p)^2-m^2,\,
                                   C_4=(k_2-p)^2-m^2,\nonumber\\ 
                                 & C_5=(k_1-k_2)^2-m^2\,.
\end{align}

The rules from \cite{Tarasov:1997kx} can be used to cancel some of scalar
products in the numerator and denominator of 
(\ref{eq:Tabxyz}), leading to
integrals with $x,y,z=0$. In particular, if we have all $n_i>0$ we are only left 
with scalar integrals without numerator. If some of $n_i$ are equal to
zero then the irreducible scalar products in the numerator, 
$(k_1 \cdot p)$ and $(k_2 \cdot p)$, cannot be canceled. Due to this, the
general form of two-loop subintegral  we need to reduce  can be cast into:
\begin{equation}
  \label{eq:Tab}
  T^{ab}_{n_1 n_2 n_3 n_4 n_5}=\int d[k_1] d[k_2]
  \frac{ (k_1 \cdot p)^a (k_2 \cdot p)^b}
  {C_1^{n_1} C_2^{n_2} C_3^{n_3} C_4^{n_4} C_5^{n_5}}\, ,
\end{equation}

The integrals~(\ref{eq:Tab}) with irreducible numerator can be reduced to a
combination of scalar integrals in shifted space-time dimension using
rules from~\cite{Tarasov:1997kx}\footnote{Alternatively reduction
  rules for the integrals with negative indices can be used and will be
  implemented in future versions of the code.}. To speed up the calculation we
prefer to use tables for such substitutions. The latter were pre-generated in advance, 
instead of generating them on the fly. The substitution rules stored in the tables distributed with the
package should be sufficient for most of practical applications and
allow to reduce integrals with $a+b \le 20$. For higher powers of
numerator, RHS of substitution rules stored in tables becomes too long
and it is more efficient to implement reduction for
integrals~\eqref{eq:Tab} with negative powers of denominators instead
of performing dimensional shifts on integrals with irreducible numerators.

As a next step we use DRR to
connect scalar integrals in shifted space-time dimension with
integrals in initial space-time dimension and reduce later to a set
of irreducible integrals. For such a purpose we follow along the
lines of~\cite{Tarasov:1997kx} and implement in the \FMFT package
a set of DRR connecting integrals with space-time dimension $d+2$ and
$d$ together with recurrence relations to reduce integrals with fixed space-time
dimension to the following master integrals:

\begin{equation}
  \label{eq:conv21}
     \begin{bmatrix}
    \texttt{F[11111]} ,\, \texttt{V[1111]} ,\, \texttt{J[211]}
    ,\, \texttt{J[111]}, \, \texttt{T2[111]}, \\
     \texttt{G[11]G[11]} ,\, \texttt{G[11]T1[1]} ,\, \texttt{T1[1]T1[1]} 
    \end{bmatrix} \otimes 
     \begin{bmatrix}
       \texttt{G[11]} \\ 
       \texttt{T1[1]} 
    \end{bmatrix}\,.
\end{equation}

In \eqref{eq:conv21} \texttt{G} and \texttt{T1} are one-loop self-energy and tadpole
integrals respectively, \texttt{T2} is the two-loop tadpole and \texttt{F,V,J} are
two-loop integrals with five, four and three lines, respectively, as
defined in \cite{Tarasov:1997kx}. All possible one- and two-loop
integrals entering convolution~\eqref{eq:conv21} are listed in~\ref{sec:ints-2l}.  

As a result of application of the above-mentioned relations, the integral in the
form~(\ref{eq:topoFGsplit}) can be represented as a sum of different
convolutions of one- and two-loop massive propagator-type
master integrals~(\ref{eq:conv21}) and a $p^2$-dependent function. This
$p^2$-dependent function has the form of a product of scalar
propagators with momentum $p^2$ and different masses, not necessary
equal to $m^2$. The masses different from $m^2$ arise from
coefficients dependent on $p^2$ and $m^2$ in front of integrals. Such a
coefficients goes to denominator when integral is substituted into
other relations. Fortunately only quadratic in $p^2$ denominators arise
during reduction of massive propagator-type integrals and
number of these new masses is fixed and the section~\ref{sec:recur-relat-tadp} is devoted to the
problem of reduction of such tadpole integrals with different masses.

\subsection{Recurrence relations for tadpoles with different masses}
\label{sec:recur-relat-tadp}

At the last stage of reduction by applying partial fractioning to $p^2$-
dependent denominators each term of the integrand of~(\ref{eq:topoFGsplit}) can be represented as a
convolution of one and two-loop integrals with fixed indices from the
set~(\ref{eq:conv21}) and a single $p^2$-dependent propagator:

\begin{equation}
  \label{eq:FGn5}
  J_i(n,m_j)=\int d[p] \frac{ F_i(p^2)G_i(p^2)}{\left(p^2-m_j^2\right)^{n}}\,.
\end{equation}

Here for each of the integrals $J_i$ the corresponding integrals $F_i(p^2)$
and $G_i(p^2)$ have fixed propagator powers given by 
combinations from~(\ref{eq:conv21}) and the mass $m_j^2$ takes one of the
possible values: $m_j^2=\{0, m^2,3 m^2, 4 m^2, 9 m^2\}$. The denominator
power $n$ can be either positive or negative, whereas for subsequent
evaluation we need to reduce it to zero or one. 

One of the possible ways to construct recurrence relations connecting
integrals in the form of~(\ref{eq:FGn5}) with different propagator
powers $n$ is to apply original Laporta ideas~\cite{Laporta:2001dd}
and derive difference equations for the integrals, in which one of
the propagator powers is treated symbolically and all others are fixed numbers.

Unfortunately, the application of Laporta reduction algorithm to the
integrals with symbolic power of one of the propagators is not a well-developed field and there is no publicly available software tools.
Due to this, we decided to use the following trick. Instead of a system of
difference equations for integrals $J_i(n,m_j)$, we construct system
of differential equations
\begin{equation}
  \label{eq:de-mx}
  \frac{\partial L_i}{\partial M^2}=A_{ik}L_k
\end{equation}
for auxiliary integrals $L_i=J_i(1,M)$ in the variable $M^2$, which
was kept as symbol during all the steps.
 
For further discussion we need to separate two cases: the first one, when
integral~(\ref{eq:FGn5}) has $n \geq 0$, and the second one, when $p^2$-
dependence is in the numerator. The second case will be considered later, but now we want to focus on the first case.

We can see that the expansion of scalar $M^2$-dependent propagator of one
of the $L_i$ integrals in a small dimensionless variable
$z=\frac{M^2-m_j^2}{m^2}$ has the following form:

\begin{equation}
  \label{eq:exp-den}
  \frac{1}{p^2-M^2}=\frac{1}{p^2-m_j^2} +
  \frac{m^2}{\left(p^2-m_j^2\right)^2} z +
  \frac{m^4}{\left(p^2-m_j^2\right)^3} z^2+\dots
\end{equation}

If we set $m_j^2$ in~(\ref{eq:exp-den}) to be equal to one of the values of
our interest, we can relate the $n$-th coefficients of $L_i$ integral
expansions in the variable $z$ with the integral $J_i(n+1,m_j)$. At the same time,
we can look for a solution of the system~(\ref{eq:de-mx}) in the form of formal series~(\ref{eq:ansatz-den}) in small variable $z$:

\begin{equation}
  \label{eq:ansatz-den}
  L_i=\sum\limits_{n=0}^{\infty}c_{i,n}z^n,\quad z=\frac{M^2-m_j^2}{m^2}.
\end{equation}

From Eqs.~(\ref{eq:exp-den}) and (\ref{eq:ansatz-den}) we can
construct the following relation~(\ref{eq:Jc-den}):

\begin{equation}
  \label{eq:Jc-den}
  J_i(n+1,m_j)=\frac{c_{i,n}}{m^{2n}},
\end{equation}
connecting the integral $J_i$ having the denominator involving $m_j^2$
in power $n$ with the  coefficients of expansion of the auxiliary
integrals $L_i$ in Taylor series in $z$. For each possible $m_j^2$
from the set $\{0, m^2,3 m^2, 4 m^2, 9 m^2\}$ expansion variable $z$
and system of equations~\eqref{eq:Jc-den} are unique. 

Substituting the ansatz (\ref{eq:ansatz-den}) into the system
(\ref{eq:de-mx}) and equating the coefficients of equal powers in $z$, we
obtain the system of difference equations in variable $n$ for the
coefficients $c_{i,n}$ and, hence, for the integrals $J_i(n,m_j)$.
The constructed system can be transformed to the triangle form and then used
for reduction of the integrals $J_i(n,m_j)$ to a set of master integrals
having the form $J_i(1,m_j)$ or $J_i(0,m_j)$. It should be noticed
that one needs to construct a separate system of recurrence relations
for all possible values of the mass $m_j^2$.

As an example, we consider recurrence relations for the integral of topology
\textbf{FG} with the following set of indices: $n_1$, $n_2$, $n_3$, $n_6$,
$n_{10} = 0$ and
$n_4$, $n_7$, $n_8$, $n_9 = 1$. This integral is a product of two
one-loop tadpoles \texttt{T1[1]} with propagators in unit power and a
two-loop vacuum integral dependent on $n$. One-loop integrals do not
contribute to difference equations and can be discarded, so it is
sufficient to write down a recurrence relation only for the two-loop part~(\ref{eq:recrel}):
\begin{align}
  \label{eq:recrel}
  \begin{tikzpicture}[baseline=(bnv1.base),scale=0.5]
    \node (bnv1) at (1,-0.3) {};
    \draw (0,0) circle (1);
    \draw[color=red] (-1,0) -- (1,0);
    \draw (0,0) node[anchor=south,shift={(0,-0.1)}]{{\tiny$n$}}; 
  \end{tikzpicture}
  = & \theta(n-3)\frac{n-d}{3(n-1)}
      \begin{tikzpicture}[baseline=(bnv1.base),scale=0.5]
        \node (bnv1) at (1,-0.3) {};
        \draw (0,0) circle (1);
        \draw[color=red] (-1,0) -- (1,0);
        \draw (0,0) node[anchor=south,shift={(0,-0.1)}]{{\tiny$n-2$}}; 
      \end{tikzpicture}
  & + & 
      \theta(n-2)\frac{d+1-2n}{3(n-1)}
      \begin{tikzpicture}[baseline=(bnv1.base),scale=0.5]
        \node (bnv1) at (1,-0.3) {};
        \draw (0,0) circle (1);
        \draw[color=red] (-1,0) -- (1,0);
        \draw (0,0) node[anchor=south,shift={(0,-0.1)}]{{\tiny$n-1$}}; 
      \end{tikzpicture}
  &
      \nonumber\\
  + & \theta(n-2)\frac{d-2}{3(n-1)}
      \begin{tikzpicture}[baseline=(bnv1.base),scale=0.5]
        \node (bnv1) at (1,-0.3) {};
        \draw (-1,0) circle (1);
        \draw[color=red] (1,0) circle (1);
        \draw (1,-1) node[anchor=south]{{\tiny$n-1$}}; 
      \end{tikzpicture}
        & + &
      \delta(n-2)\frac{2-d}{3} 
      \begin{tikzpicture}[baseline=(bnv1.base),scale=0.5]
        \node (bnv1) at (1,-0.3) {};
        \draw (-1,0) circle (1);
        \draw (1,0) circle (1);
      \end{tikzpicture}
  &
      \nonumber\\
  \begin{tikzpicture}[baseline=(bnv1.base),scale=0.5]
    \node (bnv1) at (1,-0.3) {};
    \draw[color=red] (0,0) circle (1);
    \draw (0,-1) node[anchor=south]{{\tiny$n$}}; 
  \end{tikzpicture}
  = & \theta(n-2) \frac{d+2-2n}{2(n-1)} 
      \begin{tikzpicture}[baseline=(bnv1.base),scale=0.5]
        \node (bnv1) at (1,-0.3) {};
        \draw[color=red] (0,0) circle (1);
        \draw (0,-1) node[anchor=south]{{\tiny$n-1$}}; 
      \end{tikzpicture},
       &  &  n > 1.&
\end{align}
Here $\theta(n)$ with $n \geq 0$ and $\delta(n)$ with $n=0$ are equal
to one, while for other values of $n$ both functions are equal to
zero. We can see that such ``one-dimensional'' relations affect only single
propagator power leading to small number of terms at each reduction
step and can be effectively implemented in \texttt{FORM}.


In the second case with $p^2$-dependence in the numerator
of the integral~(\ref{eq:FGn5}) it is sufficient to consider 
$m_j^2=0$ and $n \leq 0$. Such integrals with arbitrary power $n$ 
should also be reduced to the integrals $J_i(0,0)$.


We can use the same auxiliary integrals $L_i=J_i(1,M)$ dependent on
the mass $M^2$ and the system of differential equations (\ref{eq:de-mx})
as before, but now expand the scalar propagator with mass $M^2$ in the opposite
limit, i.e., in small $\bar{z}=\frac{m^2}{M^2}$:
\begin{equation}
  \label{eq:exp-num}
  \frac{1}{p^2-M^2}=- \frac{1}{m^2} \bar{z}
  - \frac{p^2}{m^4} \bar{z}^2
  - \frac{p^4}{m^6} \bar{z}^3  + \dots
\end{equation}

As before we can construct formal solution of the
system~(\ref{eq:de-mx}) in the form of series: 
\begin{equation}
  \label{eq:ansatz-num}
  L_i=\sum\limits_{n=0}^{\infty}\bar{c}_{i,n}\bar{z}^n,\quad \bar{z}=\frac{m^2}{M^2}.
\end{equation}
Then the integrals~(\ref{eq:FGn5}) with $n \leq 0$ will be related to
expansion coefficients~(\ref{eq:ansatz-num}) in $\bar{z}$
\begin{equation}
  \label{eq:J-cbar}
  J_i(-n,0)=-\bar{c}_{i,n}m^{2(n+1)}.
\end{equation}

As in the case of integrals with denominators we substitute the
ansatz~(\ref{eq:ansatz-num}) into the system~(\ref{eq:de-mx}) and
equate the coefficients in front of equal powers of $\bar{z}$. In such a
way we obtain a system of difference equations for $\bar{c}_{i,n}$, which
means that we can construct the recurrence relations for reduction of the integrals $J_i(-n,0)$
to the integrals of the type $J_i(0,0)$.

It is necessary to note that after the application of the 
recurrence relations 
to the integrals~(\ref{eq:FGn5}) with $m_j^2 \neq m^2$ the result can involve the
integrals like $J_i(1,m_j)$ with two different masses. On the other
hand, we know that if one applies traditional IBP
reduction to the fully massive four-loop tadpoles that can  be expressed
in terms of master integrals with only one mass
scale~\cite{Czakon:2004bu}. Thanks to this property all integrals
$J_i(1,m_j \neq m)$ should cancel in the final answer. Such cancellation is
a good check for correctness of the whole four-loop integrals reduction
procedure implemented in the \FMFT package.

At this step the main reduction part of \FMFT is finished and the result
is expressed in terms of symbolic expressions corresponding to master
integrals from paper~\cite{Czakon:2004bu} and coefficients dependent on $d$. For the
case of four space-time dimensions result can be expanded in $\ep$
near $d=4-2\ep$ and actual expansions for master integrals from~\cite{Czakon:2004bu}
can be substituted.

\section{Comparison with other codes and examples}
\label{sec:speed}

To estimate the \FMFT package performance and illustrate its
applicability to reduction of complicated integrals we calculate a nonplanar
integral of topology \textbf{X} (fig.~\ref{fig:fmft-X}) 
\begin{equation}
  \label{eq:Fn}
  F(n)=I(-n,1,1,1,1,1,1,1,1,1),
\end{equation}
in which the 
propagator powers are written in accordance with the auxiliary
topology~(\ref{eq:auxtopo}). The integral has a nontrivial numerator and we
compare the results of calculation for different numerator powers $n$.
The comparison was performed with the \texttt{C++} version of the
\texttt{FIRE\,5}~\cite{Smirnov:2014hma} package. Code \texttt{FIRE} is known to be
very efficient general-purpose tool for solution of the reduction problem
with many successful applications and not restricted to the reduction of
fully massive tadpoles. It can be used not only for IBP reduction with the help of Laporta
algorithm, but also in combination with the package
\texttt{LiteRed}~\cite{Lee:2012cn}. When used together
\texttt{FIRE\,5} acts as efficient tool for application of reduction
rules from resolved recurrence relations obtained by means of \texttt{LiteRed}.

\begin{table}[h]
  \centering
  \begin{tabular}{c|c|c|c|c|c|c}
    $n=$          
    & 3 & 4 & 5 & 6 & 7 & 8
    \\\hline
    \texttt{FMFT} & 0:00:11 & 0:00:27 & 0:01:55 &
                                                  0:07:35 & 0:25:31 & 01:30:31
    \\\hline
    \texttt{FIRE} & 0:01:58 & 0:09:10 & 0:28:17 &
                                                  2:16:42 & 9:19:57 & 46:42:29
  \end{tabular}
  \caption{Comparison to \FIRE, time format is \texttt{hh:mm:ss}  }
  \label{tab:fire}
\end{table}

Timing results for reduction of a single integral with \FMFT and
\texttt{FIRE\,5} are present in table~\ref{tab:fire}. For the \FMFT reduction
we use multithread version \texttt{TFORM} with eight active
workers(\texttt{-w8} option). 
Similar setup was used for   
\texttt{FIRE\,5}. It was running on eight CPU cores and in memory reduction was used (options \texttt{\#memory} and
\texttt{\#threads 8}). In addition,  the reduction rules from \texttt{LiteRed}
package were utilized. The main goal of comparison present in table~\ref{tab:fire}
is to illustrate that \texttt{FMFT} can be used for reduction of
complicated integrals as can bee seen from the time spent for integral
reduction with such efficient tool as \texttt{FIRE}.

In the listing~\ref{lst:fmft} we present simple \texttt{FORM} program  to illustrate \FMFT
usage for reduction of four-loop integrals. Integral with numerator is
defined via $p_i$ for scalar products in the numerator and $d_i=p_i^2-m^2$  for massive
denominators in correspondence to one of the top-level topologies
fig.~\ref{fig:fmft-H} and fig.~\ref{fig:fmft-X}. The main entry point is the
\texttt{fmft} routine. The result of its application is the reduction of
the initial integral to the set of master integrals identified in the
work~\cite{Czakon:2004bu} with coefficients exhibiting exact dependence
on the space-time dimension parameter $d$.
\begin{lstlisting}[language={FORM},caption={Example program},label={lst:fmft},captionpos=b]
#-
* load main library code
#include fmft.hh
* input with numerator 
L ex = p2.p3/d1/d2^2/d4/d5/d6/d7/d8/d9;
* call reduction routines
#call fmft
* expand near d=4-2*ep up to ep^1
#call exp4d(1)

b ep;
Print+s;
.end
\end{lstlisting}
By means of procedure \texttt{exp4d(n)} the result of the reduction can be expanded
in $\ep$ up to the order $\ep^n$ near $d=4-2\ep$ space-time
dimensions. Output from the program calculating integral from the
listing~\ref{lst:fmft} is the following:
\begin{lstlisting}[language={},caption={Sample output},label={lst:ex},captionpos=b]
   ex =
       + ep^-4 * ( 3/8 )

       + ep^-3 * ( 25/8 )

       + ep^-2 * ( 137/8 + 3/4*z2 - 81/4*S2 + 3/4*z3 )

       + ep^-1 * ( 363/8 - 3/2*T1ep - 1/2*z2 - 81*S2 - 3/8*z4 + 1/2*D6 - 6*z3)

       + 1/2*PR14ep0 + 1/2*PR15ep0;

       + ep * ( 1/2*Oep(1,PR14) + 1/2*Oep(1,PR15) )
\end{lstlisting}
where \texttt{z2,z3,z4} are Riemann zeta functions, \texttt{S2,T1ep}
are non zeta parts of two-loop and \texttt{D6} three-loop terms of tadpole integrals $\ep$-expansion
defined in~\cite{Steinhauser:2000ry}. Finite parts of four-loop
integrals \texttt{PR14ep0,PR15ep0} are kept as symbols and its numerical
values can be substituted from~\cite{Laporta:2002pg}. To denote
truncation of $\ep$-expansion series we use common function
\texttt{Oep} with first argument corresponding to order in $\ep$ and
second argument containing master integral name.

\section*{Acknowledgments}
\label{sec:acknowledgments}

The author would like to thank A.~Bednyakov, O.~Gituliar, B.~Kniehl, S.-O.~Moch and O.~Veretin
for stimulating discussions and for testing our program.
This work was supported in part by the German Federal Ministry for Education
and Research BMBF through Grant No.\ 05H15GUCC1 and by the German Research
Foundation DFG through the Collaborative Research Centre No.\ SFB~676
\textit{Particles, Strings and the Early Universe: the Structure of
  Matter and Space--Time}.

\appendix

\section{Two-loop massive master integrals}
\label{sec:ints-2l}
\begin{align}
  {\rm \texttt{F[11111]}}= T_{11111} & =
      \begin{tikzpicture}[baseline=(bnv1.base),scale=0.3]
        \node (bnv1) at (0,-0.15) {}; 
        \coordinate (v1) at (-3,0); 
        \coordinate (v2) at (3,0); 
        \coordinate (v3) at (0,1.75); 
        \coordinate (v4) at (0,-1.75);
        \coordinate (v5) at (4,0); 
        \draw[ndi] (v2) arc (-30:-90:3.5); 
        \draw[ndi] (v4) arc (-90:-150:3.5); 
        \draw[ndi] (v1) arc (150:90:3.5); 
        \draw[ndi] (v3) arc (90:30:3.5); 
        \draw[ndi] (v3) -- (v4); 
        \draw[ndi] (v2) -- (v5); 
        \draw[ndi] (-4,0) -- (v1);
      \end{tikzpicture}
  \\
  {\rm \texttt{V[1111]}}= T_{01111} & =  
      \begin{tikzpicture}[baseline=(bnv1.base),scale=0.3]
        \node (bnv1) at (0,-0.15) {}; 
        \coordinate (v1) at (-3,0); 
        \coordinate (v2) at (3,0); 
        \coordinate (v3) at (0,1.75); 
        \coordinate (v4) at (0,-1.75);
        \coordinate (v5) at (4,0); 
        \draw[ndi] (v2) arc (-30:-90:3.5); 
        \draw[ndi] (v4) arc (-90:-150:3.5); 
        \draw[ndi] (v1) arc (150:90:3.5); 
        \draw[ndi] (v3) arc (90:30:3.5); 
        \draw[ndi] (v1) arc (90:30:3.5); 
        \draw[ndi] (v2) -- (v5); 
        \draw[ndi] (-4,0) -- (v1);
      \end{tikzpicture}
  \\
  {\rm \texttt{J[211]}}= T_{20011} & =  
      \begin{tikzpicture}[baseline=(bnv1.base),scale=0.3]
        \node (bnv1) at (0,-0.15) {}; 
        \coordinate (v1) at (-3,0); 
        \coordinate (v2) at (3,0); 
        \coordinate (v3) at (0,1.75); 
        \coordinate (v4) at (0,-1.75);
        \coordinate (v5) at (4,0); 
        \draw[ndi] (v2) arc (-30:-90:3.5); 
        \draw[ndi] (v4) arc (-90:-150:3.5); 
        \draw[ndi] (v1) arc (150:90:3.5); 
        \draw[ndi] (v3) arc (90:30:3.5); 
        \draw[ndi] (v1) -- (v2); 
        \draw[ndi] (v2) -- (v5); 
        \draw[ndi] (-4,0) -- (v1);
        \fill (v3) circle (4pt);
      \end{tikzpicture}  
  \\
  {\rm \texttt{J[111]}}= T_{10011} & =  
    \begin{tikzpicture}[baseline=(bnv1.base),scale=0.3]
        \node (bnv1) at (0,-0.15) {}; 
        \coordinate (v1) at (-3,0); 
        \coordinate (v2) at (3,0); 
        \coordinate (v3) at (0,1.75); 
        \coordinate (v4) at (0,-1.75);
        \coordinate (v5) at (4,0); 
        \draw[ndi] (v2) arc (-30:-90:3.5); 
        \draw[ndi] (v4) arc (-90:-150:3.5); 
        \draw[ndi] (v1) arc (150:90:3.5); 
        \draw[ndi] (v3) arc (90:30:3.5); 
        \draw[ndi] (v1) -- (v2); 
        \draw[ndi] (v2) -- (v5); 
        \draw[ndi] (-4,0) -- (v1);
      \end{tikzpicture}
  \\
  {\rm \texttt{T2[111]}}= T_{11001} & =  
    \begin{tikzpicture}[baseline=(bnv1.base),scale=0.3]
        \node (bnv1) at (0,-0.15) {}; 
        \coordinate (v1) at (-3,0); 
        \coordinate (v2) at (3,0); 
        \coordinate (v3) at (0,1.75); 
        \coordinate (v4) at (0,-1.75);
        \coordinate (v0) at (0,0);
        \coordinate (v5) at (5,0); 
        \coordinate (v6) at (8,0);
        \draw[ndi] (v0) circle (1.5);
        \draw[ndi] (-4,-1.5) -- (4,-1.5); 
        \draw[ndi] (0,1.5) -- (0,-1.5); 
      \end{tikzpicture}
  \\
  {\rm \texttt{G[11]G[11]}}= T_{11110} & =  
    \begin{tikzpicture}[baseline=(bnv1.base),scale=0.3]
        \node (bnv1) at (0,-0.15) {}; 
        \coordinate (v1) at (-3,0); 
        \coordinate (v2) at (3,0); 
        \coordinate (v3) at (0,1.75); 
        \coordinate (v4) at (0,-1.75);
        \coordinate (v0) at (0,0);
        \coordinate (v5) at (4,0); 
        \draw[ndi] (v0) arc (-30:-150:1.75); 
        \draw[ndi] (v1) arc (150:30:1.75); 
        \draw[ndi] (v2) arc (-30:-150:1.75); 
        \draw[ndi] (v0) arc (150:30:1.75); 
        \draw[ndi] (v2) -- (v5); 
        \draw[ndi] (-4,0) -- (v1);
      \end{tikzpicture}
  \\
  {\rm \texttt{G[11]T1[1]}}= T_{11100} & =  
    \begin{tikzpicture}[baseline=(bnv1.base),scale=0.3]
        \node (bnv1) at (0,-0.15) {}; 
        \coordinate (v1) at (-3,0); 
        \coordinate (v2) at (3,0); 
        \coordinate (v3) at (0,1.75); 
        \coordinate (v4) at (0,-1.75);
        \coordinate (v0) at (0,0);
        \coordinate (v5) at (4,0); 
        \draw[ndi] (v0) arc (-30:-150:1.75); 
        \draw[ndi] (v1) arc (150:30:1.75); 
        \draw[ndi] (0.4,1) circle (1);
        \draw[ndi] (v0) -- (2,0); 
        \draw[ndi] (-4,0) -- (v1);
      \end{tikzpicture}
  \\
  {\rm \texttt{T1[1]T1[1]}}= T_{11000} & =  
    \begin{tikzpicture}[baseline=(bnv1.base),scale=0.3]
        \node (bnv1) at (0,-0.15) {}; 
        \coordinate (v1) at (-3,0); 
        \coordinate (v2) at (3,0); 
        \coordinate (v3) at (0,1.75); 
        \coordinate (v4) at (0,-1.75);
        \coordinate (v0) at (0,0);
        \coordinate (v5) at (4,0); 
        \draw[ndi] (0,1) circle (1);
        \draw[ndi] (0,-1) circle (1);
        \draw[ndi] (-2,0) -- (2,0);
      \end{tikzpicture}
  \end{align}

\bibliographystyle{elsarticle-num}
\bibliography{fmft}

\end{document}